# Vacuum Modified Gravity as an explanation for flat galaxy rotation curves

R. Van Nieuwenhove

*Institutt for Energiteknikk, OECD Halden Reactor Project,
Post bag 173, NO-1751 Halden, Norway
e-mail: rudivn@hrp.no*

**Abstract.**   A theory is proposed which allows explaining the observed flat galaxy rotation curves, without needing to invoke dark matter. Whereas other theories have been proposed in the past which realize the same, the present theory rests on basic physical principles, in contrast to for instance the MOND theory. The key to arrive at this new theory is to consider from the start the energy density of the vacuum. The way to calculate the effect of the corresponding vacuum pressure on a mass has previously been laid down by Van Nieuwenhove (1992). We obtain a modification of Newton's law of gravitation with some peculiar properties such as the occurrence of regions of repulsive gravity. Based on a newly derived equation of state of the vacuum, the Tully-Fisher relation is derived. The theory can make detailed predictions about galaxy rotation curves and is also able to explain to the Pioneer anomaly, the foamy distribution of galaxies and the observed accelerated expansion of the universe. A relativistic extension of the theory is included as well.

## 1. Introduction

Many physicists nowadays are convinced that some form of dark matter has to exist to explain the behavior of groups of galaxies or to explain the observed flat galaxy rotation curves. After many years of research however, this hypothetical dark matter (Saslaw 1970) has not been found and some scientists (see for instance Navarro & Van Acoleyen 2005) start doubting its existence all together. In 1983 (Milgrom 1983), the Modified Newtonian Dynamics (MOND) theory was proposed to explain the flat galaxy rotation curves without having to invoke dark matter. In 1996 (Van Nieuwenhove), the author proposed another theory to explain the flat rotation curves. This theory was based on an alternative view in which gravity results from a distortion of the quantum vacuum through its interaction with a mass (Van Nieuwenhove 1992). Using these concepts, it was shown how the existence of a large scale distortion of the vacuum energy (Van Nieuwenhove 1998) referred to as "bubble force", could explain the flat rotation curves. The required deviation of the background vacuum energy was found to be tiny (of the order of $10^{-6}$). Nevertheless, it was shown that this resulted in a dramatic modification of the motion of stars within the "vacuum bubble". The weakness of this theory was however that the vacuum bubble was put in by hand. Secondly, it was found that the properties of this vacuum bubble were somehow coupled to the mass of the galaxy, while no explanation was provided at that time. Thirdly, the shape of the vacuum bubble was not calculated but was instead based on an intelligent guess. Finally, no new predictions could be made to confirm the proposed theory. In this paper, an extended version of this original theory has been elaborated which eliminates all of the above objections.

## 2. Different aspects of vacuum energy

In the Einstein field equations (Misner, Thorn & Wheeler 1973), the cosmological constant appears on the left side in

$$R_{\mu\nu} - \frac{1}{2} \cdot R \cdot g_{\mu\nu} + \Lambda \cdot g_{\mu\nu} = \frac{8\pi G}{c^4} \cdot T\mu\nu \qquad (1)$$

where R and g pertain to the structure of spacetime and T pertains to matter. The cosmological constant has the same effect as an intrinsic energy density of the vacuum ($\rho_{vac}$). However, it needs not to have a counterpart in quantum physics. It serves the function of an integration constant in that its presence does not violate the conservation properties of the Einsteinian tensor. Nevertheless, the relation between $\Lambda$ and the energy density of the vacuum is often made. The equation of state for this vacuum is given by p = - ρ in which p is the pressure. So, a positive energy density corresponds to a negative pressure and this was the equation of state which was used in (Van Nieuwenhove 1992). In 1998 (Caldwell & Dave & Steinhardt), another form of energy, called quintessence (or "fifth element"), was proposed to explain the observations of an accelerating rate of cosmic expansion. Unlike $\Lambda$, quintessence can vary over space and has an equation of state given by p = w ρ where w is less than -1/3. Whereas $\Lambda$ is part of the geometric tensor (left side in equation (1)), quintessence enters on the right hand side of the equation, thought of as a scalar field. Based on these new concepts, it seems better to replace the terminology of "vacuum energy", as used in (Van Nieuwenhove 1992, 1998) by "quintessence" since we are relying on energy of the vacuum which is varying over space. In this paper, we will however further use the terminology of the "vacuum energy" in the loose sense.

Whereas the effect of vacuum energy is normally evaluated by means of the Einstein Field Equations, a method has been proposed in (Van Nieuwenhove 1992) to evaluate its effect on the motion of matter by means of a simple classical description. The method consists in representing a mass (m) by a volume through the relation

$$V_m = \frac{mc^2}{\rho_V} \qquad (2)$$

in which $\rho_V$ corresponds to the (background) energy density of the vacuum. The next step consists in evaluating the force F on this mass by

$$F = V_m \cdot \frac{dp}{dr} \qquad (3)$$

in which r is the radial coordinate and where we use the convention that F is positive when attractive. Equation (3) means that the mass experiences a net force when the pressure of the vacuum is not balanced out. So, in this theory, a truly constant vacuum energy density (such as described by Λ) can not induce a force, however large the energy density may be (Van Nieuwenhove 1992).

As soon as one starts changing the vacuum properties ("deforming") one induces however gradients such that the deformed vacuum itself starts to act as a source of gravitation. But how can we evaluate this? The gravitational force is, non-relativistically given by Newton's law

$$F = \frac{G \cdot m_1 \cdot m_2}{r^2} \tag{4}$$

Since energy and mass can transform according to $E = m \cdot c^2$, one can consider a thought experiment where one of the masses (say $m_1$) in equation (4) is transformed somehow in vacuum energy (or quintessence). Then we assume that equation (4) is still valid. In this case we can write that:

$$m_1 = \frac{4\pi}{c^2} \cdot \int_0^{r_1} \rho(r) \cdot r^2 \cdot dr \tag{5}$$

where $\rho(r)$ is the spherically symmetric vacuum energy density and the integral extends from r = 0 to the position ($r_1$) of the test mass $m_2$. It is important here to consider for $\rho(r)$ only that part which is varying over space. This is because a constant contribution can not induce a gravitational force, as explained before.

### 3. Vacuum modified gravity

Consider next a mass M and let us deduce the total force between this mass M and a test mass m. Since the energy in the vacuum (at least the varying part) contributes to the gravitational force, we can write

$$F = \frac{G \cdot m \cdot (M + M_V)}{r_1^2} \tag{6}$$

where $M_V$ is given by

$$M_V = \frac{4\pi}{c^2} \cdot \int_0^{r_1} \rho(r) \cdot r^2 \cdot dr \tag{7}$$

Inserting equation (7) into equation (6) and using equations (2,3) and the equation of state p = - αρ (with α positive) and equating the force given by equation(6) to the force given by equation (3) results in

$$\int_0^{r_1} \rho(r) r^2 dr = -a \cdot r^2 \frac{d\rho(r)}{dr} - b \tag{8}$$

where

$$a = \frac{\alpha c^4}{4\pi G \rho_V} \tag{9}$$

and

$$b = \frac{c^2 M}{4\pi} \tag{10}$$

After some lengthy algebra, one finds that

$$\rho(r) = \frac{1}{\left(r/\sqrt{a}\right)} \cdot \left[ p \cdot \cos\left(\frac{r}{\sqrt{a}}\right) + q \cdot \sin\left(\frac{r}{\sqrt{a}}\right) \right] \tag{11}$$

where

$$p = \frac{b}{a^{3/2}} \tag{12}$$

and q is an arbitrary constant. The gravitational force can then be expressed as (see equation (3))

$$F = -\alpha c^2 \cdot \frac{m}{\rho_V} \cdot \frac{d\rho(r)}{dr} \tag{13}$$

Inserting equation (11) into equation (13), one obtains

$$F = -\alpha c^2 \cdot \frac{m}{\rho_V} \cdot \left\{ \frac{-\sqrt{a}}{r^2} \cdot \left[ p \cdot \cos\left(\frac{r}{\sqrt{a}}\right) + q \cdot \sin\left(\frac{r}{\sqrt{a}}\right) \right] + \frac{1}{r} \cdot \left[ -p \cdot \sin\left(\frac{r}{\sqrt{a}}\right) + q \cdot \cos\left(\frac{r}{\sqrt{a}}\right) \right] \right\} \tag{14}$$

For small values of r ($r \ll \sqrt{a}$), this can be approximated by

$$F = \alpha c^2 \cdot \frac{m}{\rho_V} \cdot \frac{\sqrt{a} \cdot p}{r^2} \tag{15}$$

It can easily be verified that this is nothing else than Newton's law (by inserting a and p). For larger values of r, the form of the gravitational force becomes however considerably more complicated. It will be shown further that q = p.

We can work at that

$$p = \frac{b}{a^{3/2}} = \frac{MG^{3/2}(4\pi)^{1/2} \rho_V^{3/2}}{\alpha^{3/2} c^4} \tag{16}$$

Further we will define R (with a dimension of length) to be

$$R = \sqrt{a} = \frac{\alpha^{1/2} c^2}{\sqrt{4\pi G \rho_V}} \qquad (17)$$

Introducing further the dimensionless ratio x = r/R, we can write the force as

$$F = \frac{GmM}{R^2} \left[ \frac{1}{x^2}[\cos(x) + \sin(x)] - \frac{1}{x}[-\sin(x) + \cos(x)] \right] \qquad (18)$$

So, by means of some very basic physical principles and a minimum of assumptions we have arrived at a modified Newtonian description of the gravitational force and propose to designate this approach by Vacuum Modified Gravity (VMG). At this stage, the value of R is not yet defined. This information will however be obtained from galaxy rotation curves, as explained before.

It is interesting to see (equation (18)) that the force is composed of two parts. We therefore define the following functions

$$f_1(x) = \frac{1}{x^2}(\cos(x) + \sin(x)) \qquad (19)$$

$$f_2(x) = \frac{1}{x}(-\sin(x) + \cos(x)) \qquad (20)$$

$$f_t(x) = f_1(x) + f_2(x) \qquad (21)$$

$$f_N(x) = \frac{1}{x^2} \qquad (22)$$

$$f_p(x) = -\frac{1}{x}(\cos(x) + \sin(x)) \qquad (23)$$

The function $f_t(x)$ is proportional to the total force and $f_N$ is the Newtonian force. The function $f_p$ is proportional to the vacuum pressure.

The behavior of these functions (except of f1(x)) is shown in Fig. 1. The function $f_1(x)$ still has a close resemblance to the Newtonian dependence given by $f_N(x)$ but the function $f_2(x)$ is however of a different character. In fact, one can see that this corresponds to the "bubble force" which was originally introduced in (Van Nieuwenhove 1998), without derivation. The total force $f_t(x)$ has an interesting behavior. At x = 3.660 the force changes from being attractive to being repulsive and becomes attractive again at
x = 6.925. It oscillates then further from positive to negative up to infinity. The bubble force starts out negative (from r = 0) and becomes positive at x = 0.785 and switches back to negative at x = 3.927 and oscillates further up to infinity. So, because of the peculiar action of the vacuum, every mass will induce, at some large radius, regions of repulsive force.

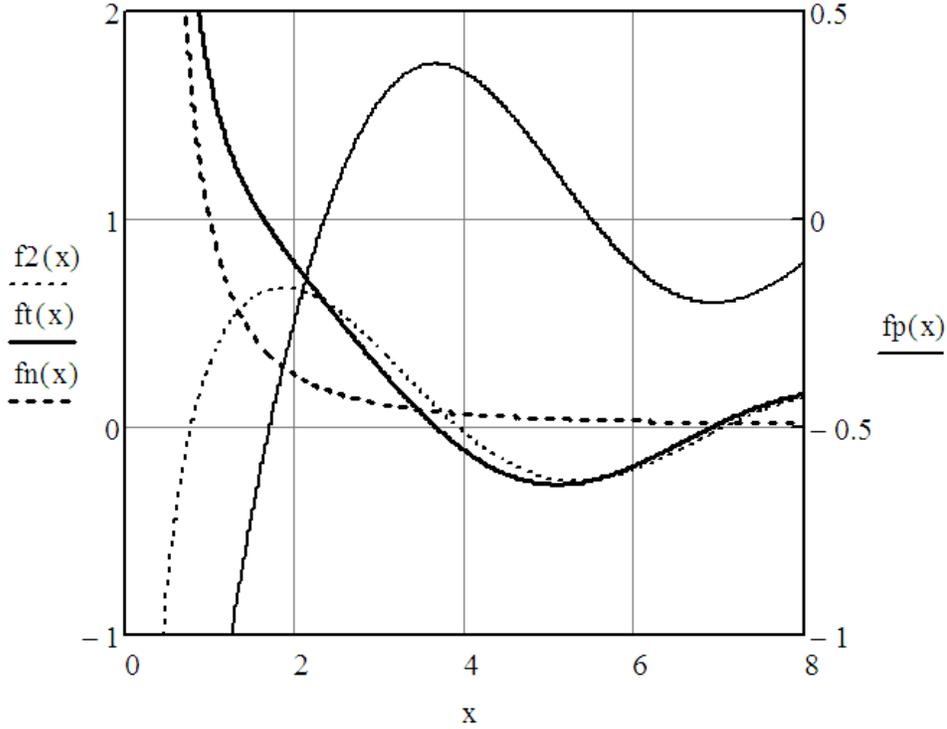

**Figure 1.** Radial dependencies of the normalized forces f2(x), ft(x), fn(x), as defined in equations 20-23. Note that x is given by x = r /R. The profile of the vacuum pressure is also shown (right scale). The behavior of the vacuum modified gravity force ft(x) is markedly different from the Newtonian gravity force fn(x) when x is not very small.

### 4. Galaxy rotation curves

Before applying the modified gravity equations to a galaxy, we need also to derive the equations inside the bulge of the galaxy. We will also neglect completely the mass in the disk and in the halo. So, we are dealing here with a highly simplified model. Its purpose is to demonstrate some qualitative features rather than to model a realistic galaxy.

Let us denote the bulge radius by $r_b$ and the normalized bulge radius by $x_b = r_b / R$. The total mass contained in the bulge is M. The mass inside a sphere of radius $r_1$ can then be written as

$$M_{r_1} = \frac{r_1^3}{r_b^3} \cdot M \qquad (24)$$

The equivalent mass of the vacuum energy, contained in a sphere of radius $r_1$ is given by equation (7). The equation to solve for $\rho(r)$ then becomes

$$\int_0^{r_1} \rho(r) r^2 dr = -a \cdot r^2 \frac{d\rho(r)}{dr}\bigg|_{r=r_1} - \frac{z}{3} \cdot r_1^3 \qquad (25)$$

in which a has the same meaning as before (see equation (9)) and z is given by

$$z = \frac{3Mc^2}{r_b^3 \, 4\pi} \tag{26}$$

To solve for ρ(r), we expand ρ(r) in powers of r, determine the corresponding coefficients by solving equation (25) and reassemble the function again in a concise form. Then we obtain:

$$\rho(r) = -z + \frac{h}{(r/R)} \cdot \sin\left(\frac{r}{R}\right) \tag{27}$$

in which $x = \frac{r}{\sqrt{a}} = \frac{r}{R}$ and h is an arbitrary constant. The resulting force is then given by:

$$F = \frac{GmM}{R^2} \cdot d \cdot \left[\frac{\cos(x)}{x} - \frac{\sin(x)}{x^2}\right] \tag{28}$$

The relation between d and h is given by:

$$h = d \cdot \frac{Mc^2}{4\pi R^3} \tag{30}$$

The free parameter h is obtained by requiring that *both* the vacuum energy density and its derivative are matching each other at the transition from the inside to the outside solution(at x = $x_b$). These two conditions allow obtaining both h and q.
After some algebra, one finds then that

$$q = -z \cdot (x_b \cdot \cos(x_b) - \sin(x_b)) = -\frac{3}{x_b^3}(x_b \cdot \cos(x_b) - \sin(x_b)) \cdot p \tag{31}$$

Further, one finds that

$$h = \frac{\sin(x_b) + q \cdot \cos(x_b) + x_b \cdot z}{\sin(x_b)} \tag{32}$$

Since $x_b$ is small (< 0.1), one can make an expansion of the cosine and sine terms in Eq.(31):

$$q \approx -\frac{3}{x_b^3} \cdot \left[x_b\left(1 - \frac{x_b^2}{2!}\right) - x_b + \frac{x_b^3}{3!}\right] \cdot p \tag{33}$$

leading thus to the very important result that

$$q = p = \frac{Mc^2}{4\pi R^3} \tag{34}$$

Further, it can be shown that the vacuum energy density at the origin is to a reasonable approximation (valid for small values of $x_b$) equal to

$$\rho(0) \approx \frac{1.55}{x_b} \cdot p \approx \frac{1.55 \cdot Mc^2}{4\pi \cdot r_b \cdot R^2} \tag{35}$$

Equating the centripetal force to the gravitational force provides an equation for the rotation velocity (v):

$$V_{in} = \sqrt{d \cdot \left(\frac{\sin(x)}{x} - \cos(x)\right)} \tag{36}$$

in which $V_{in}$ is the normalized velocity given by:

$$V_{in} = \frac{v}{\sqrt{\frac{GM}{R}}} \tag{37}$$

The radial dependence of the force is shown in Fig.2.
Next, we consider the region outside of the bulge of the galaxy.
Equating again the gravitational force (equation (18)) to the centripetal force results in the following velocity profile:

$$V_{out} = \sqrt{\frac{1}{x} \cdot (\cos(x) + \sin(x)) + (\sin(x) - \cos(x))} \tag{38}$$

in which

$$V_{out} = \frac{v_{out}}{\sqrt{\frac{GM}{R}}} \tag{39}$$

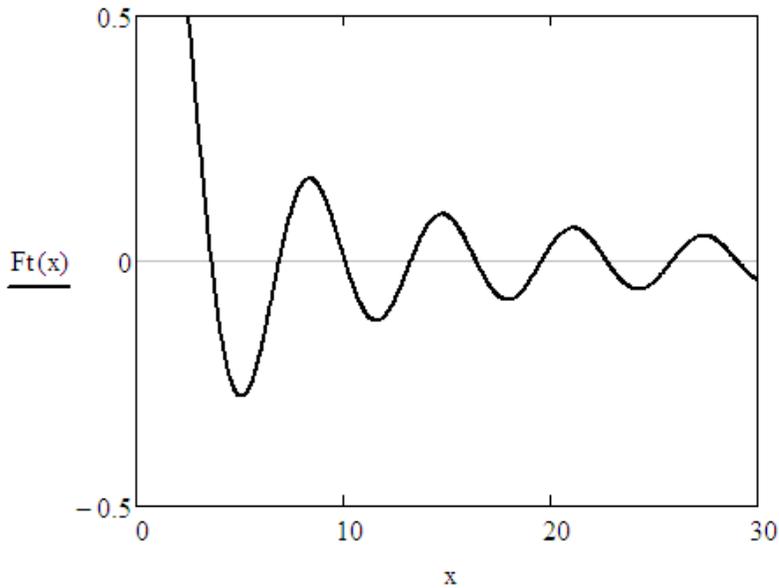

**Figure 2**. Radial dependence of the vacuum modified gravity force showing the remarkable behavior that the force oscillates between being attractive and being repulsive.

The complete velocity profile is shown in Fig. 3. As can be seen from Fig. 3, we obtain a plateau region in velocity. No fitting parameters were obtained for obtaining the flat velocity region. It just follows from consistently taking into account the vacuum energy density. From the calculated velocity profile, one can observe a number of interesting features. The velocity does not remain constant up to infinity (as in the MOND model) but goes to zero at (approximately) x = 3.66. The plateau velocity comes out to be

$$v_{pl} = \sqrt{\frac{GM}{R}} \cdot 1.28 \tag{40}$$

R.Van Nieuwenhove           arXiv:0712.1110 [physics.gen-ph] 7 Dec 2007

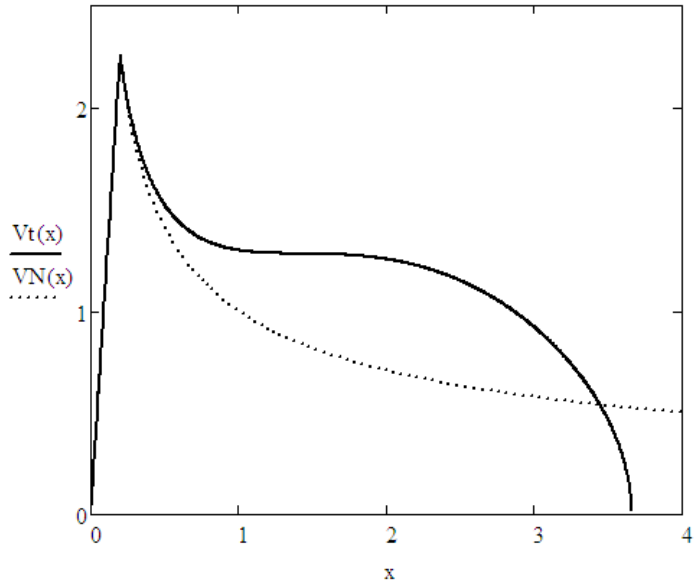

**Figure 3**. Predicted normalized velocity profile Vt(x) for an idealized galaxy (consisting of a central bulge only) as based on vacuum modified gravity and comparison to the Keplerian velocity profile VN(x) where x is given by $x = r/R$. While a region of constant velocity ($V_{pl}$) is observed, the velocity drops to zero at $x = 3.66$.

Inside the bulge region, the velocity profile is very close to that obtained by using Newton's law. From Fig. 2 one sees that the force oscillates between positive and negative when moving further out, as discussed earlier. As a result, matter which was originally present in between the zero crossings will unavoidably move towards the location of these zero crossings and accumulate there. The present analysis concentrates only on the field structure of one galaxy. In the presence of several galaxies, the wavy vacuum perturbations will overlap and interfere, creating complicated patters. This could possibly explain the fact that galaxies form a bubble-like or "foamy" distribution in space with galaxies lying along the surfaces of the bubbles (called "voids") and nothing in the centers of the bubbles (Gregory & Thompson 1978).

On a cosmological scale, the present model can even explain the observed accelerated expansion of the universe which started about 5 billion years ago. Indeed, from equation (17), one sees that if one assumes a very small value of $\rho_V$, one can obtain an R value comparable to the size of the universe. When the universe expands, eventually the gravitational force will turn into a repulsive force (see figure 2) and hence will decelerate the expansion. This deceleration will however only apply beyond a given radius.

The attentive reader will have noticed that equation (40) is in contradiction with the Tully Fisher relation (Tully & Fisher 1977), *unless* one requires that

$$R = \beta \cdot \sqrt{M} \qquad (41)$$

In this case one obtains that $v_{pl}^4 \approx M$.

Inserting equation (41) into equation (35) results in

$$\rho(0) = \frac{2.41 \cdot 10^{-4}}{r_b(kpar\sec)} \cdot \frac{J}{m^3}$$ and thus independent of the mass of the galaxy.

Using galaxy data, as provided in (Sanders & McGaugh 2002), one finds that the value of β should be about 1.22 $m/\sqrt{kg}$, though it is not our aim here to pinpoint the most exact value. Let us now consider as an example our own galaxy. We assume a mass of $4 \times 10^{41}$ kg and using equation (41) one finds that R = 25 kpc and by means of equation (40) we find $v_{pl}$ = 238 km/s. Based on the value of R (25 kpc), the velocity profile is predicted to fall to zero at r = 3.66 x 25 kpc = 91.5 kpc. The velocity is however expected to start dropping significantly at 60 kpc. The above mentioned values could of course be different when assuming a more realistic mass distribution.

Inserting equation (41) into equation (17), one finds that:

$$\rho_V = \frac{\alpha c^4}{4\pi\beta^2 MG} \quad (42)$$

Our own solar system could represent a very small perturbation on top of the vacuum deviation (global $\rho_V$) of our galaxy as a whole. In this respect, it is found that the Pioneer anomaly can be reproduced perfectly when assuming a value of β given by β = 0.2 $m/\sqrt{kg}$. Calculating the acceleration (from equation (18)) and subtracting from it the acceleration as obtained by Newton's law, one obtains an additional, almost constant, sunward acceleration of 8 x 10$^{-10}$ m/s$^2$. Of course, β has been chosen here to fit the observation and this value is different from the one used previously (1.22 $m/\sqrt{kg}$) to describe the galaxy rotation curves. Nevertheless, it is remarkable that reducing the scale length by 8 orders of magnitude (galaxy to solar system), β remains of the same order. According to equation (41), very large mass distributions will lead to oscillating force fields having a very large scale length. This could result in voids with extremely large dimensions.

## 5. Discussion

From Eq.(42) one finds that $\rho_V$ is inversely proportional to M. The only way to make $\rho_V$ a constant, independent of M is to assume that α is proportional to M. Then we could write α as:

$$\alpha = k \cdot \frac{4\pi \cdot G \cdot \rho_V}{c^4} \cdot M \quad (43)$$

in which k is the proportionality constant. In this case, one finds that β (see Eq.(41)) is related to k by $k = \beta^2$. The constant k has then dimension $m^2/kg$. Coincidentally, the value of k (1.488 $m^2/kg$) is roughly equal to the "surface" of the universe (calculated as $4\pi \cdot R_u^2$) divided by the mass of the universe. Indeed, using the presently accepted (most likely) values of $R_u = 3.7 \cdot 10^{26} m$ and $\rho_M \approx 9.9 \cdot 10^{-27} kg/m^3$ (calculating M by $M = \rho_M \cdot \frac{4\pi \cdot R_u^3}{3}$), one finds $k = \frac{3}{R_u \cdot \rho_M} = 0.819 \cdot \frac{m^2}{kg}$ which is of the same order as the previously assumed value of k (1.488 $m^2/kg$). So, k obtained in this way is a factor 1.816 too small. If one would have assumed relation (43) from the start, the Tully Fisher relation would have followed automatically. One may wonder why α would have to be proportional to the mass M (Eq. (43)). A simple derivation based on energy conservation in an expanding universe (Appendix A) shows that this is indeed the case. From Appendix A (Eq.(12)), one finds that α is given by

$$\alpha = \frac{1}{2} \cdot \frac{\frac{GM}{c^2}}{R_u} \cdot \frac{\rho_{e,u}}{\rho_V} \tag{44}$$

This implies that we have actually *derived* the Tully-Fisher relation.
From this expression, one can derive (based on Eq.(17)) that

$$k = \beta^2 = \frac{1}{8 \cdot \pi} \cdot \frac{c^2}{R_u} \cdot \frac{\rho_{e,u}}{\rho_V^2} \tag{45}$$

Equating this to $1.816 \cdot \frac{3}{R_u \cdot \rho_M}$ (see above), results in

$$\rho_{e,u} = \sqrt{137} \cdot \rho_V \tag{46}$$

The factor 137 is (accidentally?) equal to one over the fine structure constant $\alpha_{fine} = \frac{e^2}{\hbar c 4\pi\varepsilon_0}$. Since the energy density of the vacuum is related to quantum fluctuations (through Heisenberg's uncertainty relation, characterized by the constant $\hbar$), it can be expected that the fine structure constant enters a more complete theory of gravitation.

## 6. Relativistic vacuum modified gravity

In general relativity, the line element summarizing the Schwarzschild geometry is given by ($c \neq 1$ units):

$$ds^2 = -\left(1 - \frac{2GM}{c^2 r}\right) \cdot (cdt)^2 + \left(1 - \frac{2GM}{c^2 r}\right)^{-1} \cdot (dr)^2 + r^2 \cdot \left(d\theta^2 + \sin^2\theta \cdot d\phi^2\right) \qquad (47)$$

To obtain the relativistic expression for vacuum modified gravity, one has to replace the potential -GM/r by the potential given below:

$$\Phi = -\frac{MG}{r} \cdot \left(\cos\left(\frac{r}{R}\right) + \sin\left(\frac{r}{R}\right)\right) \qquad (48)$$

in which R is given by equation (41).

# 7. Conclusions

A modified, relativistic theory of gravitation has been proposed in which the vacuum energy density plays a key role. This theory builds further on a theory which has been proposed already in 1992 and applied further in 1996 (Van Nieuwenhove). At each point in space, Newton's law holds locally if one takes into account the additional energy in the vacuum contained in an enclosed sphere up to that point. Globally, one obtains however a different dependence of force on distance and the exact equation for this has been derived. It is found that any mass will, at some determined regions, induce repulsive forces.

This modification of Newton's law has a "free" scale parameter R which depends on the environment (or overall mass distribution) in which the mass is located. It has been shown that this theory (designated by Vacuum Modified Gravity) naturally explains the observed flat galaxy rotation curves without the need to invoke dark matter. The Tully Fisher relation implied that $R = \beta \cdot \sqrt{M}$. This relation was found to be consistent with a newly derived equation of state of the vacuum. To describe correctly the galaxy rotation curves, it was found that $\beta = 1.22\ m/\sqrt{kg}$. The factor $k = \beta^2$ was found to be roughly equal to the surface of the universe divided by its mass. It was found that the relation between the (mass) energy density of the universe is related to the (quantum) energy density by $\rho_{e,u} = \sqrt{137} \cdot \rho_V$. Based on a newly derived equation of state of the vacuum, the Tully-Fisher relation is derived. Unlike the MOND theory, the present theory predicts that the rotation curves eventually drop of sharply to zero velocity. The proposed theory also explains the Pioneer anomaly, the foamy distribution of galaxies and the accelerated expansion of the universe. In addition, large voids are explained as a result of large mass distributions creating large regions of repulsive force (antigravity).

## References

Anderson, J. D., Laing, P.A., Lau, E.L., Liu, A. S., Nieto, M. M., Turyshev, S. G., 1998, *Phys. Rev. Lett*. **81**, 2858.
Caldwell, R. R., Dave, R., Steinhardt, P. J. 1998, *Phys. Rev. Lett.,* **80**, 1582.
Gregory, S.A., Thompson, L.A., 1978, *ApJ*, **222**, 784.
Milgrom, M. , 1983, *ApJ*, **270**, 365.
Misner, C. W., Thorn, K. S., Wheeler, J. H. 1973, Gravitation, W.H. Freeman and Company, San Francisco, California.
Navarro, I., Van Acoleyen, K., 2005, *Phys. Lett.*, **B622**, 1.
Sanders, R. H., McGaugh, S. S. 2002, *Annu. Rev. Astron. Astrophys.,* **40**, 263.
Saslaw, W. C. 1970, *Astrophys. J., ***160**, 11.
Tully, R. B., Fisher, J. R. 1971, *Astron. Astrophys.*, **54**, 661.
Van Nieuwenhove, R. 1992, *Europhys. Lett.,* **17**, 1.
Van Nieuwenhove, R. 1998, *Astronomical and Astroph. Tr.,* **16**, 37.
Van Nieuwenhove, R. 2007, *Concepts of Phys.*, **4**, 645.

# Appendix A
**The equation of state of the vacuum**

If the vacuum energy density ($\rho_V$) would be an absolute constant, independent of space, time and matter, one can show that

$$p = -\rho_V \tag{1}$$

The relation (1) can also be deduced by a means of a thought experiment in which a cylinder, filled by vacuum energy, is pulled out (through a piston). When the volume of the piston increases, the total enclosed energy increases since $\rho_V$ is assumed to be independent of everything. If the additional volume created is $\Delta V$, the additional energy is $\rho_V \cdot \Delta V$. The work done by the piston is $W = p \cdot \Delta V$ in which p is the vacuum pressure. Energy conservation requires then that $p = -\rho_V$. This well-known classical argument is however based entirely on the assumption that $\rho_V$ is independent of everything. If one associates $\rho_V$ with the quantum fluctuations of the vacuum however, such an assumption can no longer be maintained. It is well-known that matter is in constant interaction with the surrounding vacuum and therefore $\rho_V$ must be influenced by it. It was this recognition that led to VMG in the first place. In the following, we therefore derive a more realistic equation of state of the vacuum.

Consider a galaxy with mass M (our reference galaxy) in an expanding universe, characterized by a homogenous mass density $\rho_{m,U}$ (the index m indicates mass and U universe). Just as for the previous example with the moving piston, we investigate next the change in energy due to the expansion of the universe and impose conservation of energy to arrive at the equation of state of the vacuum. Here we assume that $\rho_V$ behaves in a more realistic way in the sense that it will decrease due to the expansion, such that the overall energy remains constant. Therefore we do not need to consider $\rho_V$ further in our energy balance. When the universe expands, the overall gravitational potential energy of all the masses in the galaxy, relative to our reference galaxy (with mass M) will increase. This increase in energy is calculated next.

Consider a spherical shell at a distance r from our reference galaxy. The mass inside this shell is given by

$$dm = \rho_{m,U} \cdot 4\pi \cdot r^2 dr \tag{2}$$

And the potential energy relative to the reference galaxy is given by

$$dE_p = -G \cdot M \cdot \rho_{m,U} \cdot 4\pi r \cdot dr \tag{3}$$

Integrating Eq.(3) from r = 0 to r = $R_u$ (radius of the universe), leads to

$$E_p = -G \cdot M \cdot 4\pi \cdot \rho_{m,U} \cdot \frac{R_U^2}{2} \tag{4}$$

Further we have that

$$\rho_{m,U} = \frac{A}{R_U^3} \tag{5}$$

In which A is a constant.

Inserting Eq.(5) into Eq.(4) results in

$$E_p = -G \cdot M \cdot 4\pi \cdot \frac{A}{2 \cdot R_u} \tag{6}$$

The variation in potential when the radius $R_u$ is increased to $R_u + dR$ is then given by

$$dE_p = G \cdot M \cdot 4\pi \cdot \frac{A}{2} \cdot \frac{1}{R_u^2} = \frac{4\pi GM}{2} \cdot \rho_{m,U} \cdot R_u \tag{7}$$

Further we have that

$$\rho_{e,u} = \rho_{m,u} \cdot c^2 \tag{8}$$

In which $\rho_{e,u}$ is the energy density of the universe (corresponding to its mass density).
Inserting Eq.(8) into Eq.(7) results in

$$dE_p = 2\pi \cdot \frac{GM}{c^2} \cdot \rho_{m,U} \cdot R_u \cdot dR_u \tag{9}$$

The work done by the outer surface (S) of the universe (in analogy to our piston) is given by

$$dW = p \cdot S \cdot dR_u \tag{10}$$

In which

$$p = -\alpha \cdot \rho_V \tag{11}$$

in which α is a dimensionless factor.

Since $dE_p + dW = 0$, one can work out that (combining Eq.9,10,11)

$$\alpha = \frac{1}{2} \cdot \frac{\frac{GM}{c^2}}{R_u} \cdot \frac{\rho_{e,u}}{\rho_V} \tag{12}$$

Resulting in

$$p = -\frac{1}{2} \cdot \frac{\frac{GM}{c^2}}{R_u} \cdot \frac{\rho_{e,u}}{\rho_V} \cdot \rho \tag{13}$$

So, here we obtain finally the remarkable result that the proportionality factor α (in the equation of state Eq.(11)) is proportional to the mass M. So, the factor α is not an absolute factor; it depends on the chosen reference galaxy (and so of the chosen mass M). Therefore, the equation of state is different for different observers. Further, it should be remarked that the above analysis is only valid over sufficiently large scale lengths.

We will assume further that $\rho_{e,u}$ is linearly proportional to $\rho_V$.